\documentclass[journal=ancac3, manuscript=article, layout=traditional]{achemso} 
\usepackage[utf8]{inputenc}
\usepackage[T1]{fontenc}
\usepackage[english]{babel}
\usepackage{microtype}

\usepackage{libertine, libertinust1math}
\usepackage{xcolor}
\usepackage{siunitx, physics}
\usepackage[version=4]{mhchem}
\usepackage[margin=10pt, format=hang, font=small, labelfont=bf,
labelsep=space]{caption}

\newcommand{\fc}[1]{\textbf{({#1})}}
\newcommand{\rev}{\textcolor{black}}

\title{Enhancing vibrational light-matter coupling strength beyond the molecular concentration limit using plasmonic arrays}
\author{Manuel Hertzog}
\affiliation{Department of Chemistry and Molecular Biology,
  University of Gothenburg, Kemigården 4, 412 96, Gothenburg, Sweden}
\author{Battulga Munkhbat}
\author{Denis G. Baranov}
\author{Timur O. Shegai}
\affiliation{Department of Physics, Chalmers University of Technology,
  412 96, Gothenburg, Sweden}
\email{timurs@chalmers.se}
\author{Karl Börjesson}
\affiliation{Department of Chemistry and Molecular Biology,
  University of Gothenburg, Kemigården 4, 412 96, Gothenburg, Sweden}
\email{karl.borjesson@gu.se}


\begin{document}


\begin{abstract}
  Vibrational strong coupling is emerging as a promising tool to modify
  molecular properties, by making use of hybrid light-matter states known as
  polaritons. Fabry-\rev{Perot} cavities filled with organic molecules are typically
  used, and the molecular concentration limits the maximum reachable coupling
  strength. Developing methods to increase the coupling strength beyond the
  molecular concentration limit are highly desirable. In this letter, we
  investigate the effect of adding a gold nanorod array into a cavity
  containing pure organic molecules, using FT-IR microscopy and numerical
  modeling. Incorporation of the plasmonic nanorod array, that acts as
  artificial molecules, leads to an order of magnitude increase in the total
  coupling strength for the cavity filled with organic molecules. Additionally,
  we observe a significant narrowing of the plasmon linewidth inside the
  cavity. We anticipate that these results will be a step forward in exploring
  vibropolaritonic chemistry and may be used in plasmon based bio-sensors.
\end{abstract}

\section{Introduction}
Strong light-matter coupling has attracted considerable attention in the past
couple of years due to the potential applications it offers in physical and
chemical sciences.\cite{Ebbesen2016,Ribeiro2018,Flick2018,hertzog2019strong}
For example, strong coupling of organic molecules has been shown to modify the
rate of a photoisomerization reaction,\cite{Schwartz2011, Hutchison2012}
increase electronic transport,\cite{Orgiu2015} and expand the length scale of
Förster energy transfer.\cite{Coles2014,Zhong2016,Zhong2017} Not to mention
other effects of strong coupling such as selective manipulation of excited
states,\cite{Stranius2018} suppression of photo-oxidation \cite{Munkhbat2018}
or reducing photodegradation in polymers.\cite{Peters2019} Recently,
vibrational strong coupling has come into focus as a promising physical tool to
control molecular properties. Since the first experimental evidences of
vibrational strong coupling (VSC) in both
solid\cite{Shalabney2015,Long2014,Simpkins2015} and liquid
states,\cite{George2015} the field has expanded
considerably,\cite{Hertzog2020,Menghrajani2019,Dunkelberger2019,Pietron2019,Xiang2019,Menghrajani2019a,Menghrajani2020,Takele2020}
and it has been shown to alter reaction
kinetics,\cite{Lather2019,Hirai2020,Pang2020,Thomas2020} control reaction
selectivity,\cite{Thomas2019} allow for intermolecular energy
transfer,\cite{Xiang2020} and modification of enzyme
activity.\cite{Vergauwe2019} Recent progresses in chemical reactions making use
of polaritons were summarized by Hirai et al.\cite{Hirai2020a} In order to
significantly impact chemical reactivity, theoretical investigations has
demonstrated that a high coupling strength is
required\cite{Galego2019,Hiura2019} and a recent experimental study has shown a
non-linear relationship between the coupling strength and thermodynamics of a
chemical reaction.\cite{Thomas2020}

Strong light-matter coupling is achieved by interfacing molecules with confined
electromagnetic field of resonant cavities tuned to a molecular
transition. When a molecular vibrational transition is in the strong coupling
regime, two new hybrid states, known as polaritons, are
formed,\cite{Khitrova2006} separated in energy by the so-called Rabi splitting
$\hbar\Omega_{\text{R}}$. Traditionally, polaritons have been engineered with
the use of planar cavities, such as Fabry-\rev{Perot} resonators confining the
electromagnetic fields between two mirrors.\cite{Shalabney2015} Reaching large
coupling strengths usually requires saturating the cavity volume with the
molecular material, and additionally aligning the transition dipole moments
with the cavity vacuum field. \cite{hertzog2017voltage,Berghuis2020,Roux2020}

Plasmonics offers an alternative route to strong coupling by confining light
down to subwavelength scales with the use of metallic nanoparticles and
nanocavities.\cite{Savasta2010,Manjavacas2011,Schlather2013} Only a tight
region of space around the metallic cavity needs to be filled with molecules in
order to form polaritons.\cite{Chikkaraddy2016} Plasmon resonances of
nanoparticles are tunable from the UV to the IR range, and can be used for
molecular sensing in the IR range due to nanoscale mode
volumes.\cite{Rodrigo2015,Limaj2016,Singh2016,Leitis2019} However, to improve
the sensitivity, new approaches for narrowing the plasmon linewidth are
desirable.

The magnitude of the Rabi splitting for a given molecular transition is
proportional to the square root of the molecular concentration and the filling
factor. Furthermore, the maximal achievable magnitude of the Rabi splitting in
the case of saturated mode volume is ultimately bounded by the bulk Rabi
splitting $\hbar\Omega_{\text{R}} \sim \sqrt{f}$ (with $f$ being the transition
oscillator strength \cite{Novotny2006}), which is independent on the cavity
type.\cite{Toermae2014,Pino2015,Baranov2017} Therefore, new approaches are
required to increase the Rabi splitting in order to maximize the effect of the
confined electromagnetic field onto molecules, as shown by theoretical
studies.\cite{Climent2019,Galego2019,Triana2020}

In this Letter, we utilize a hierarchical coupling between a Fabry-\rev{Perot}
cavity, a vibrational absorption band of an organic molecule, and a localized
surface plasmon resonance in the mid-IR regime to go beyond the Rabi splitting
imposed by the maximal concentration limit \cite{Bisht2018}. First we show that
the coupling of the plasmon and the FP cavity results in an order of magnitude
decrease in plasmon's linewidth, an observation rationalized by reduced
radiative losses from the plasmon in the cavity \cite{Baranov2020}. Then, by
using numerical and analytical modelling in conjunction to our experimental
data, we show a five to nine-fold increase in the total coupling strength,
indicating that the plasmon act as an artificial molecule that increases the
molecular coupling strength.

\section{Results \& discussions}

\begin{figure}[H]
  \centering \includegraphics[width=165mm]{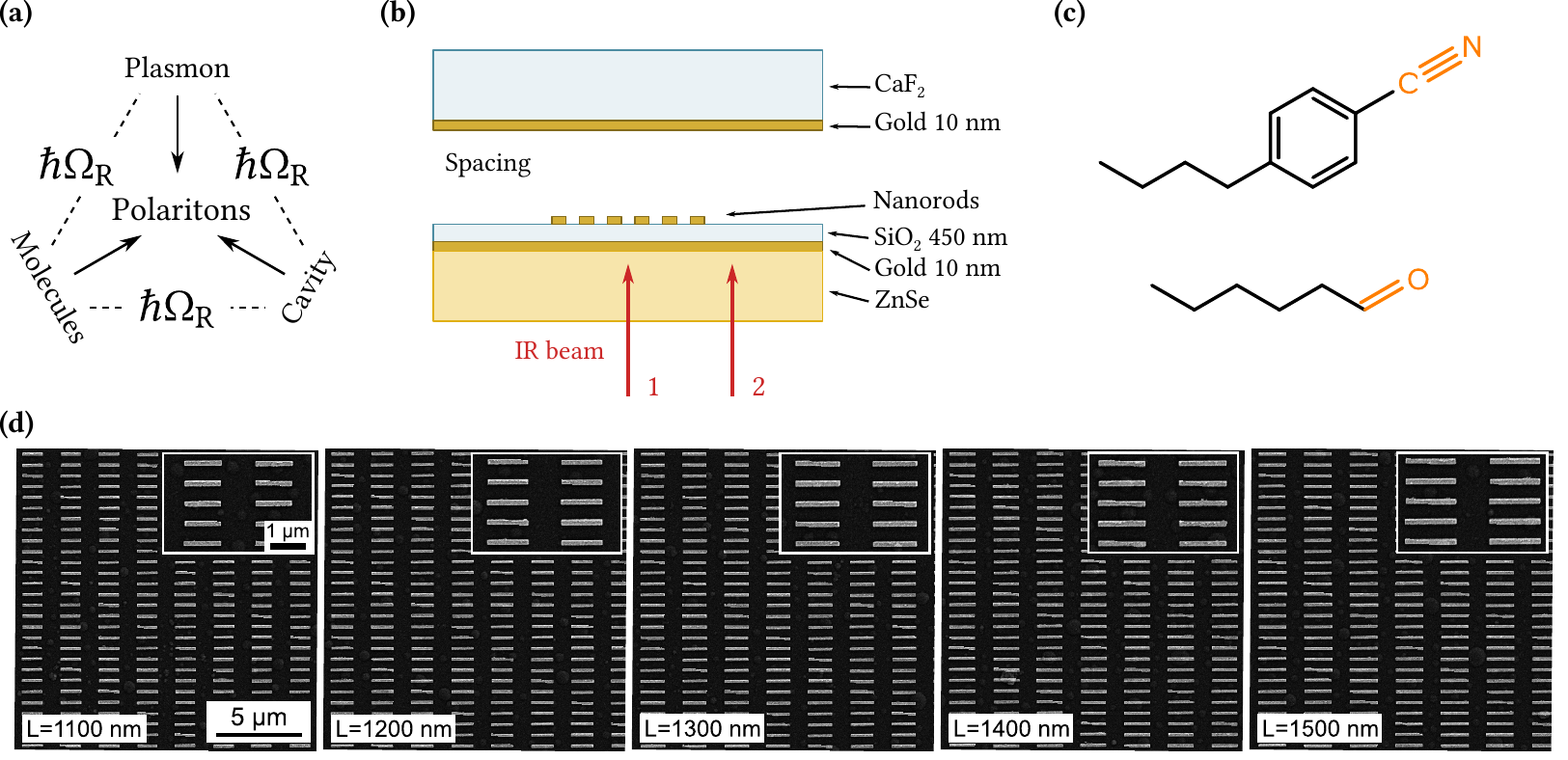}
  \caption{\fc{a} Conceptual diagram of the system in this study. \fc{b}
    Schematic view of the Fabry-\rev{Perot} cavity. Fourier transform infrared
    spectroscopy microscopy was used to either probe the small area where gold
    nanorods were deposited (beam 1) or where no rods where present (beam
    2). \fc{c} Molecular structure of 4-butylbenzonitrile (top) and hexanal
    (bottom). Highlighted in orange are the functional groups responsible for
    the vibrational band coupled to the FP cavity. \fc{d} SEM picture of the
    gold nanorods deposited on a glass substrate.}
  \label{fig:fig_1}
\end{figure}
\begin{figure}[H]
  \centering \includegraphics[width=165mm]{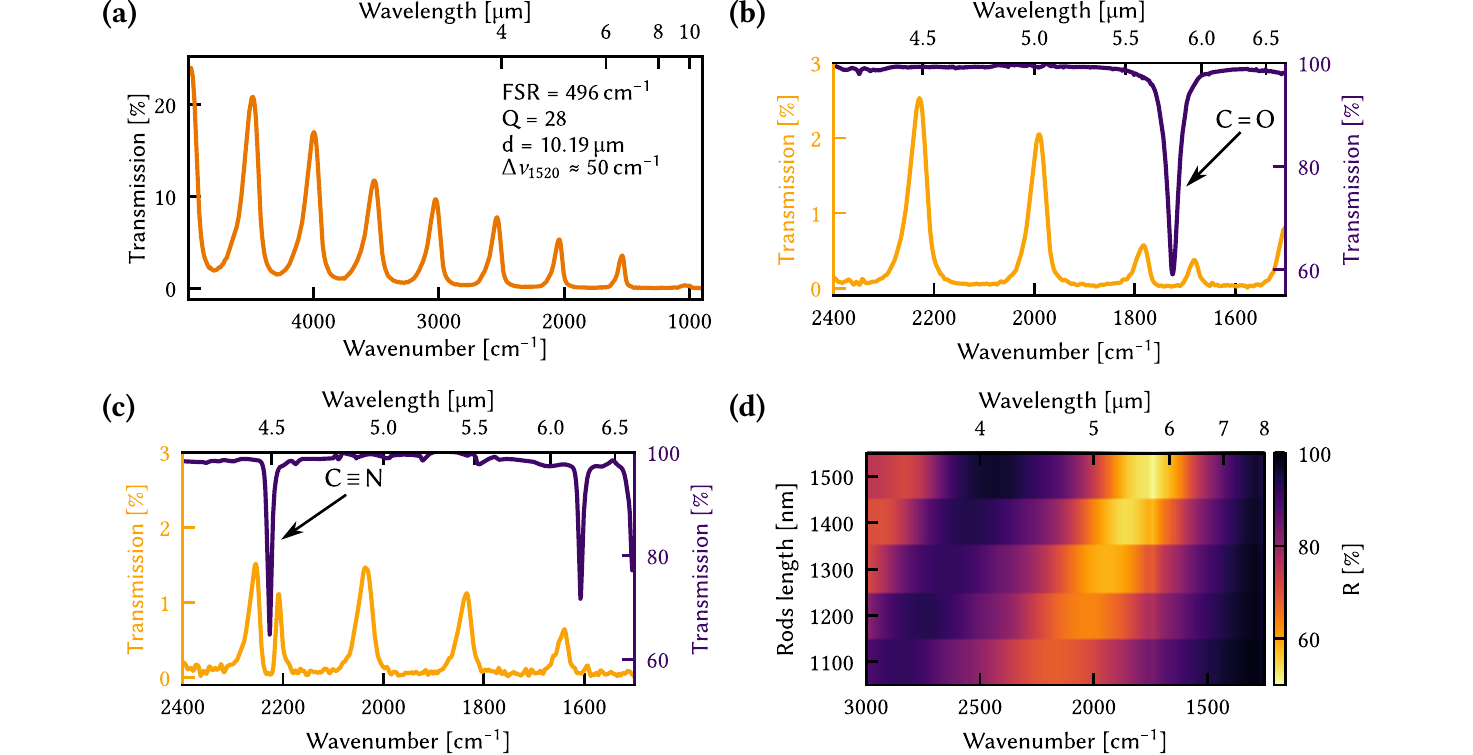}
  \caption{\fc{a} Transmission spectrum of an empty Fabry-\rev{Perot} cavity. \fc{b}
    ATR spectrum of hexanal (purple) showing the \ce{C=O} absorption band
    around \SI{1724}{\per\cm}, and transmission spectrum of the FP cavity
    filled with hexanal (orange). The value of $\hbar\Omega_{\text{R}}$ is
    \SI{101}{\per\cm}.  \fc{c} ATR spectrum of 4-butylbenzonitrile (purple)
    showing the \ce{C\bond{3}N} absorption band around \SI{2225}{\per\cm}, and
    transmission spectrum of the FP cavity filled with 4-butylbenzonitrile
    (orange). The value of $\hbar\Omega_{\text{R}}$ is \SI{46}{\per\cm}. \fc{d}
    Reflection map of the five different gold nanorod arrays.}
  \label{fig:fig_2}
\end{figure}

In this study, we report a method to increase the coupling strength above the
limit of $\sqrt{C}$ using a hybrid system composed of a Fabry-\rev{Perot} (FP)
cavity, an organic molecule, and a localized surface plasmon, in a fashion
similar to the one introduced by Bisht et al.\cite{Bisht2018} using
two-dimensional transition metal dichalcogenides in the visible regime. All
three entities are tuned to the same resonance frequency, thereby coupled
amidst themselves, creating hybrid polaritons (Figure \ref{fig:fig_1}a). The
Fabry-\rev{Perot} cavities used in the following experiments are composed of
IR-transparent substrates (\ce{CaF2} and \ce{ZnSe}) coated with \SI{10}{\nm} of
gold (Figure \ref{fig:fig_1}b; see SI for experimental methods). The physical distance between the gold mirrors,
ranging from 11 to \SI{16}{\um}, was controlled using a polymer spacer. The
cavities were designed with two inlets allowing to inject liquids. The quality
factor of an empty cavity was 28, and the free spectral range 496 cm$^{-1}$
(Figure \ref{fig:fig_2}a).  We choose hexanal and 4-butylbenzonitrile as the
organic molecules for this study (Figure \ref{fig:fig_1}c). Both are liquids at
room temperature and were processed in neat form. The molecular vibrations of
interest are the \ce{C=O} stretching mode of hexanal and the \ce{C#N}
stretching mode of 4-butylbenzonitrile. These are strong absorption bands
centered around \SI{1724}{\per\cm} and \SI{2225}{\per\cm}, with full width at
half maxima (FWHM) of 27 and \SI{10}{\per\cm}, respectively
(Figure~\ref{fig:fig_2}b and \ref{fig:fig_2}c). The surface plasmon was
provided by gold nanorod arrays (size ca. \SI{100}{\um^2}), which where
deposited on top a spacer composed of \SI{450}{\nm} of \ce{SiO2} to minimize
interference from the gold mirror (Figure \ref{fig:fig_1}b). Five different
arrays were made, all having equal thickness and spacing but lengths ranging
from 1100 to \SI{1500}{\nm}, which gave plasmon resonances covering the energy
of both two molecular vibrations and the Fabry-\rev{Perot} mode (\rev{Figure
\ref{fig:fig_2}d}). The plasmon resonances were broad (FWHM = 621--1120
\si{\per\cm}) due to radiative losses. This is an intrinsic feature of plasmon
arrays in the mid-IR, hampering its use.

We will first describe the coupling between the cavity and the two \rev{organic compounds}, then the cavity and plasmon, and lastly, the complete coupled system with the
FP cavity, the plasmon and the \rev{organic molecules}. Figure~\ref{fig:fig_2}b and c
show a clear normal mode splitting of the vibrational absorption band of the
two molecules when placed inside the FP cavity at resonant conditions. This
indicates strong coupling between the cavity mode and the carbonyl group of
hexanal (Figure~\ref{fig:fig_2}b), and the cavity mode and the nitrile group of
4-butylbenzonitrile (Figure~\ref{fig:fig_2}c). The resulting formation of
vibro-polaritons gives a measured Rabi splitting of \SI{101}{\per\cm} and
\SI{46}{\per\cm} for the \ce{C=O} mode and the \ce{C#N} mode with the cavity,
respectively. The Rabi splittings are larger than both the FWHM of the bare
molecular vibrations and the cavity mode, which constitutes further evidence
that our system is indeed in the strong-coupling regime. Furthermore the ratio
between the coupling strength and the bare transition energy of the vibrations
are 2.93\% for the \ce{C=O} band and 1.03\% for the \ce{C#N} band.

\begin{figure}[H]
  \centering \includegraphics[width=165mm]{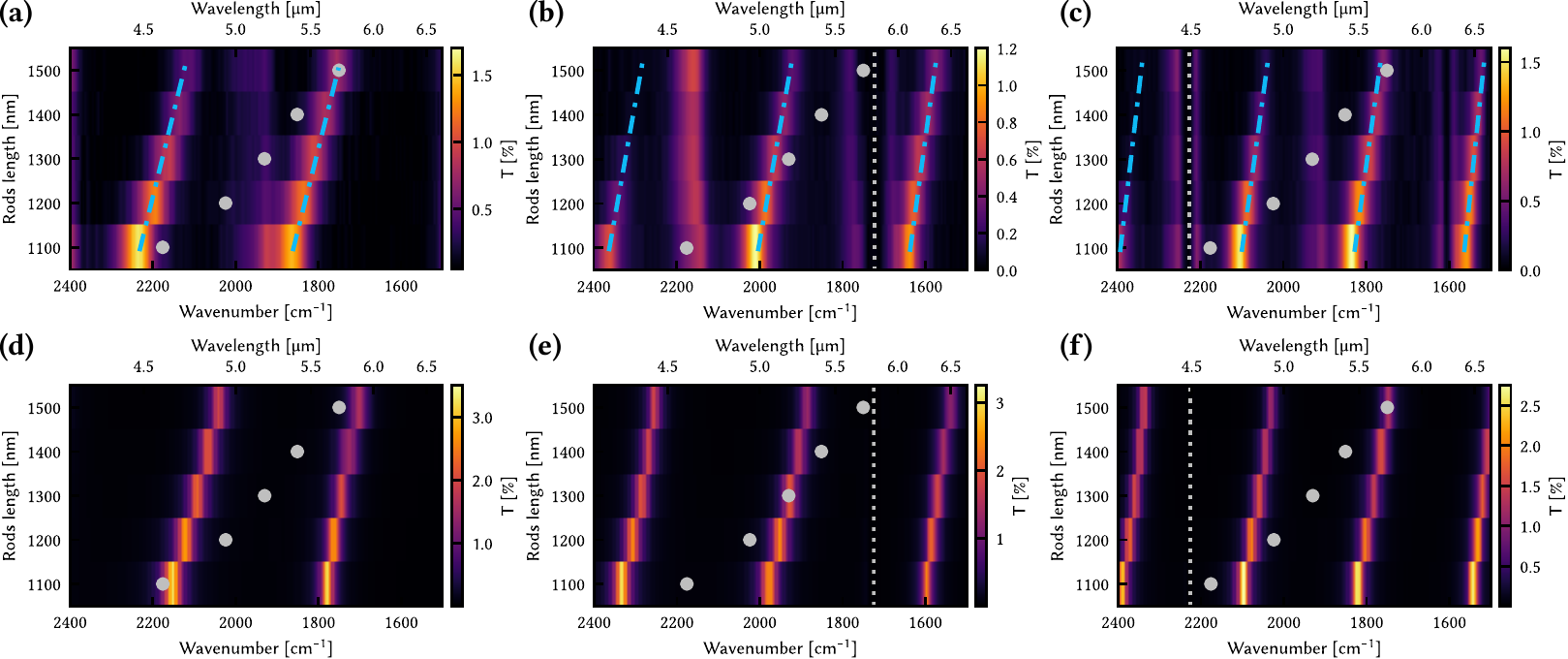}
  \caption{\fc{a)-(c} Transmission maps of the gold rods inside the Fabry-\rev{Perot}
    cavity, containing air, hexanal, or 4-butylbenzonitrile, respectively. All
    three were measured with a polarizer along the long axis of the rods. The
    blue dash-dot lines highlight the newly formed polaritonic states, the
    residual vertical modes are artifact due to non-ideal polarization
    alignment. \fc{d)-(f} Simulated spectra of the same systems \rev{(with the cavity thicknesses of \SI{10.8}{\um}, \SI{8.5}{\um} and \SI{10.4}{\um}, respectively)}. The gray
    dashed line indicates the absorption band of interest of the molecules, and
    the gray dots indicate the plasmon absorption maximum.}
  \label{fig:fig_3}
\end{figure}

Let us now consider the plasmonic arrays inside the FP cavity. With increasing
rod length, the plasmon resonance shifts to lower energies
(Figure~\ref{fig:fig_2}d), and the nanorod length therefore provides a way of controlling the cavity-plasmon detuning. Moreover, due to intrinsic anisotropy
of the rods, the microcavity-plasmon polaritons are observed only along the long
axis of the rods, while along the short axis, the bare cavity is
recovered. This polariton anisotropy can be probed using polarization-resolved
transmission spectroscopy. As shown in Figure~\ref{fig:fig_3}a, 
\rev{the two original FP cavity modes that can be observed in the given spectral range couple to the plasmonic array mode, forming exactly $2+1=3$ microcavity-plasmon polaritonic states, denoted with dot-dashed lines. These new modes exhibit almost linear dispersive behavior because of the underlying dispersion of the plasmonic mode with the nanorod length.}  

Furthermore, we observe an order of magnitude decrease in the linewidth of the
resulting polariton compared to the bare plasmon. Similar narrowing was
observed in our previous experiments with plasmonic arrays ultra-strongly
coupled with FP cavities in the visible to mid-IR range.\cite{Baranov2020} The
linewidth of closely packed metallic nanorods in the IR region is dominated by
radiative losses; but when placed inside a closed cavity, the radiation from
the nanorods does not instantaneously leave the cavity, instead bouncing
between the mirrors and thus reducing the total resonance linewidth.
Consequently, the FWHM of the plasmon polaritons drops to ca. \SI{70}{\per\cm}
for all the rods' lengths, as a result of significantly suppressed radiative
damping. Observing that the FWHM of the FP modes are $84\pm7.3$ \si{\per\cm},
the linewidth of the cavity is the main limitation of the linewidth of the
plasmon polariton (Table S2).

By adding molecules into the cavity, a third resonant component is introduced
to the hybrid system. The hexanal \ce{C=O} band is on resonance with the
longest nanorods (Figure~\ref{fig:fig_3}b), whereas the 4-butylbenzonitrile
\ce{C#N} band is on resonance with the shortest nanorods
(Figure~\ref{fig:fig_3}c). This time, more FP modes are present in the same spectral range due to the larger background refractive index of the molecular solution compared to air. \rev{Dot-dashed lines indicate the dispersions of the resulting cavity-plasmon-molecule polaritonic modes, whereas the dashed vertical lines indicate the respective molecular resonances. In the next section, we will uncover the polaritonic origin of these linearly-dispersing modes with the use of the analytical Hamiltonian formalism.}
Furthermore, the polaritonic linewidths are again
reduced by the cavity mirrors, showing that the FP cavity mode limits the
polariton linewidth irrespectively of the medium inside the cavity.

In order to gain further understanding of the experimental results, we
performed numerical modelling of the coupled systems \rev{with the use of the finite-difference time domain (FDTD) method in a commercial software (Lumerical}
Figure~\ref{fig:fig_3}d-f and Figures S1-3). Simulated normal-incidence transmission spectra demonstrate a good agreement with the experimental spectra for both plasmonic structures, and composite plasmon-molecule ones. Specifically, the
observed polaritonic modes linearly disperse with the rods length, however, the
dispersion is less pronounced than for uncoupled rods (gray dots). This is a
consequence of intermixing between highly dispersive plasmonic modes and
non-dispersive FP modes (all arrays were placed in the same FP cavity).


\rev{
Generally, a system of $N$ distinct oscillators produces $N$ polaritonic modes upon coupling (as long as the system is not exactly at an exceptional point). In our case, we start with $M$ bare cavity modes, the plasmonic array mode, and the molecular resonance, so that one could expect emergence of $M+2$ new eigenstates. However, the simulated spectra strongly suggest that the molecules and the plasmons join in a single effective oscillator, which adds only one additional polaritonic mode to the system, as one can clearly see from the transmission dispersions in Fig. 3 (aside from the parasitic uncoupled FP modes).
This confirms that even a system with slightly detuned plasmons and molecules does behave as a system with a single additional compound plasmonic-molecular oscillator.}

Next, to ensure that the observed dispersions are a result of the interaction
between the three components in the hybrid system, transmission spectra of the
system with a polarizer perpendicular to the nanorods' long axis were
measured. The results are shown in Figure~S5 (numerical
modeling in Figure S1). As expected, when the contribution of the plasmonic
array is removed by the polarizer, the transmission spectra are the same as
when probing the system beside the array (Figure~\ref{fig:fig_2}b and
c). Likewise, the values of $\hbar\Omega_{\text{R}}$ with a perpendicular
polarizer are \SI{46}{\per\cm} and \SI{101}{\per\cm} for the \ce{C#N} and
\ce{C=O} vibrations, respectively, which are the same values as we observed
when when probing next to the plasmonic array.

\section{Theoretical analysis}
In order to extract coupling strengths and confirm that the plasmonic array
acts as an artificial molecule to enhance the total coupling strength of the
system, we turn to theoretical analysis of the experimental results. This
analysis is essentially based on the coupled harmonic oscillator algebra in the
simplest possible implementation. Furthermore, this analysis requires several
rather crude assumptions, which may be false in general, but are satisfactory
for the goal of extracting the collective coupling constants.

In view of the above remark, we describe the cavity by a set of $M$ orthogonal
Fabry-\rev{Perot} eigenmodes, each coupling to the molecular resonance with a certain coupling constant. As we are far from the ultrastrong coupling regime in this case, the interacting system can be described by multimode coupled-harmonic oscillator Hamiltonian, which under the rotating wave approximation takes the form (see Methods):
\begin{equation}
  \hat{H}_{mol}=\sum_{m=1}^M {\hbar \omega_m \hat a_m ^{\dagger} \hat{a}_m } + \hbar \omega_0 \hat b ^{\dagger} \hat{b} + 
  \sum_{m=1}^M \hbar g_m (\hat a ^{\dagger}_m \hat{b} + \hat{a}_m \hat b ^{\dagger})
\end{equation}
where $\hat a$ and $\hat b$ are the annihilation operators of the $m$-th cavity
mode and that of the molecular resonance, respectively, and $g_m$ is the
coupling constant.

The coupled Fabry-\rev{Perot} system exhibits transmission peaks at its polaritonic
resonances, corresponding to the eigenvalues of the Hamiltonian (Eq. 1).
Therefore, we estimate the cavity-molecule coupling strength by fitting the
energies of the transmission peaks by the eigenvalues of Hamiltonian Eq. 1
accounting for $M=20$ lowest cavity modes (see Methods for the details of the fitting procedure). 
Since in the multi-mode system the coupling strength is
dependent on the FP mode order and the frequency, to make a reasonable
comparison we estimate the coupling strength at zero detuning: that is, with
the particular FP mode, (near-)resonant with the molecular transition. 

By fitting the Hamiltonian eigenvalues to the positions of measured transmission peaks (see Figure S6), we obtain the hexanal-cavity system zero-detuning coupling strength (with the 7-th FP mode of a \SI{14.25}{\um} thick cavity) of about \SI{41}{\per\cm}. Remarkably, this value is close to the bulk polariton
coupling strength of hexanal
$g_{bulk}=\omega_p\sqrt{f/\varepsilon_\infty}/2 \approx 49$ \si{\per\cm}, which
describes the photon-molecule interaction strength in an unbounded homogeneous
molecular medium.\cite{Hopfield1958} Similarly, for the
4-butylbenzonitrile-cavity system, we obtain the zero-detuning coupling
strength (with the 11-th mode of a \SI{16.15}{\um} thick cavity) of about
\SI{18}{\per\cm}, which is also comparable to the respective bulk coupling
strength of 4-butylbenzonitrile (\SI{21}{\per\cm}).

\begin{figure}
  \centering \includegraphics[width=165mm]{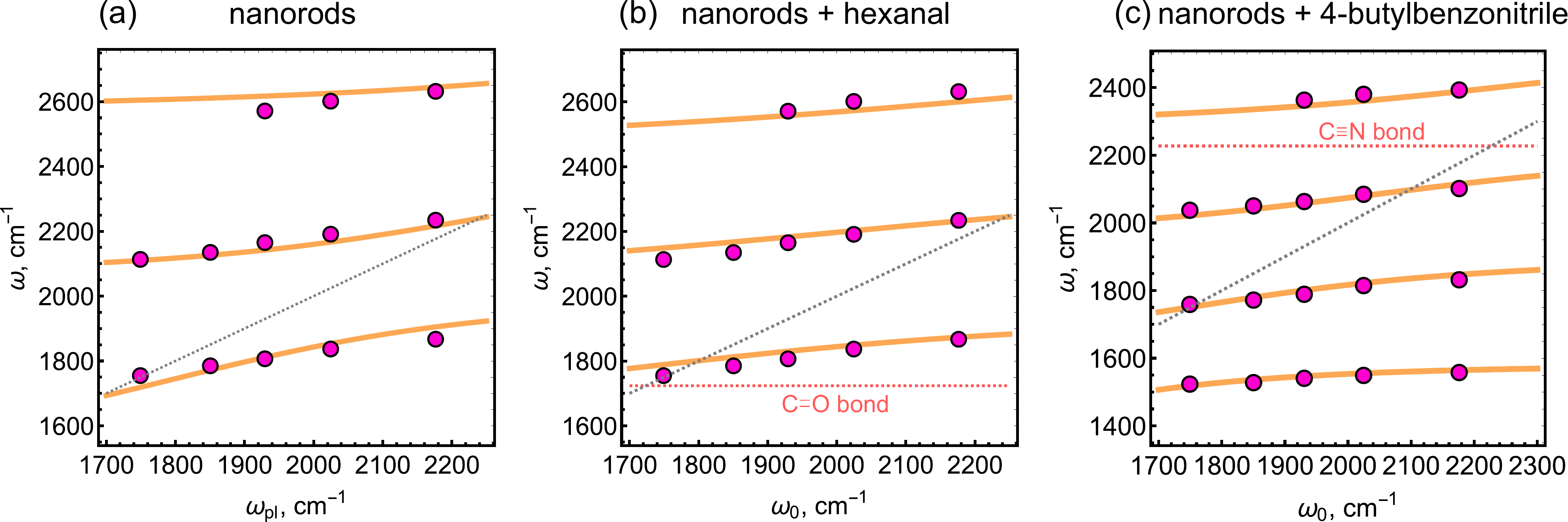}
  \caption{Hamiltonian analysis of the experimental data. \fc{a} Fitting of the
    measured transmission peaks of the coupled nanorod-cavity system (circles)
    with eigenvalues of the multimode JC Hamiltonian (lines). \fc{b} The same
    as (a) for the hybrid hexanal-nanorod system. \fc{c} The same as (a) for
    the hybrid 4-butylbenzonitrile-nanorod system.}
  \label{fig:fig_4_fits}
\end{figure}

In a similar way, we analyze the cavity-nanorod system by replacing the single collective molecular resonance with a single mode plasmonic mode (see Methods). 
By fitting the measured dispersion of transmission peaks 
with the Hamiltonian eigenvalues, we obtain the energy spectrum of the coupled system, Figure 4(a).
\rev{For as long as the plasmonic array is located in a specific horizontal plane $z=z_0$ of the cavity, the coupling strength strongly disperses with the mode order $m$ as the result of alternating electric field nodes and anti-nodes of the cavity's standing wave (see Methods).
Considering the 4-th FP mode of a \SI{9.7}{\um} thick cavity resonant at about \SI{2000}{\per\cm} (which is close to both molecular resonances),} we obtain the nanorod array coupling strength to that cavity mode of about \SI{172}{\per\cm}. 
\rev{To help appreciate the appearance of anti-crossings in this multi-mode system, we present the resulting dispersion of the Hamiltonian eigenvalues in the full spectral range in Fig. S7, which clearly demonstrate a set of anti-crossings between the polaritonic states.}

Finally, we switch to the complete cavity-plasmon-molecule systems. The key idea of our experiment is to show that plasmonic nanorods can act as artificial
molecules with the oscillator strength significantly exceeding molecular ones,
joining the real molecules in a single collective oscillator, and thus boosting
the coupling strength and Rabi splitting. Therefore, we model the full three-component structures with the same Jaynes-Cummings Hamiltonian, in which the plasmon-molecule hybrid is described as a single collective oscillator (see Methods).
By fitting the measured dispersion of transmission peaks of the hybrid
plasmon-hexanal-cavity system with the Hamiltonian eigenvalues, Figure 4b, we
obtain a resonant coupling strength (with the 5-th FP mode of a \SI{8.0}{\um}
thick cavity and $z=500$ nm) of about \SI{220}{\per\cm}.  Despite the number of
assumptions made in the analysis, the Hamiltonian fit shows a good agreement
with measured transmission peaks.  Similarly, for the hybrid
plasmon-4-butylbenzonitrile-cavity system we obtain a resonant coupling
strength of \SI{202}{\per\cm} (with the 8-th FP mode of a \SI{10.1}{\um} thick
cavity), Figure 4c. \rev{Full spectral range dispersions obtained for these samples, shown in Fig. S7, again clearly demonstrate a set of anti-crossings between polaritonic states, which within the experimentally accessible domain appear as straight lines.}

The simple analysis performed above suggests that the molecules and the
nanorods act indeed as a collective oscillator with the effective coupling
strength \rev{very} approximately given by the sum of the two individual coupling
strengths, $g_{eff} \sim g_{pl} + g_{mol}$. Thus, the presence of the plasmon
boosts the effective coupling strength of 4-butylbenzonitrile and hexanal with
the cavity mode by 10 times and 5 times, respectively.  The results of our
experiment and its subsequent analysis suggest that using the plasmon resonance
with large oscillator strength indeed allows to boost the effective coupling
strength beyond the bulk limit bounded by the molecular concentration. Of
course, this addition of the plasmonic meta-atoms does not modify the molecular
oscillator strength per se, but rather modifies the effective polaritonic
spectrum of the hybrid system in the vibrational strong coupling regime. This
in turn may potentially affect chemical reactions whose rate was claimed to
depend on the vacuum Rabi splitting in the recent literature. \cite{Thomas2019}

\section{Conclusion}

In summary, our hybrid Fabry-\rev{Perot} cavities show that the addition of a
plasmonic array to the standard molecular vibro-polaritonic system increases
the total coupling strength by almost an order of magnitude for a nitrile
absorption band and five times for a carbonyl absorption band. Increasing the
coupling strength beyond the molecular concentration limit, dismantle the
crucial obstacle for reaching the ultra-strong coupling regime using organic
molecules. Furthermore, precisely controlling the coupling strength, not only
with the molecular concentration, but also with the density of the plasmonic
array, allows molecules at small concentrations to reach the strong coupling
limit \rev{in a complementary fashion than the cooperating vibrational strong coupling reported by Lather et al.\cite{Lather2019}} One can in a sense view the plasmonic array as a form of "catalyst",
that enables any on-resonance molecular transition, regardless of molecular
concentration and transition dipole moment strength, to reach the strong
coupling regime. Furthermore, the cavity reduces the radiative damping from the
plasmon, sharpening the polariton linewidth with more than an order of
magnitude.  Together with the spectral tuning ability, such sharp linewidths
may allow for mode-selective chemical sensing in the mid-IR. The approach
described here is not limited to infrared transitions, but can also be
transposed to electronic transitions. For these reasons, we suggest that our
hybrid system will be an ideal platform to explore the promising potential of
polaritonic chemistry, the ultra-strong coupling regime, as well as provide an
approach to mode-selective mid-IR sensing.

\begin{acknowledgement}
  B.M., D.G.B. and T.O.S. acknowledge financial support from the Swedish Research
  Council (under VR Miljö project, grant No: 2016-06059) and the Knut and Alice
  Wallenberg Foundation. K.B. and M.H. acknowledge financial support from the European Research council (ERC-2017-StG-757733) and the Knut and Alice Wallenberg Foundation (KAW 2017.0192).
\end{acknowledgement}

\begin{suppinfo}
  The supporting information contains the following sections.
  \begin{itemize}
    \item S1: Methods
    \item S2: Additional figures

  \end{itemize}
\end{suppinfo}
\bibliography{references_nano_letters}

\end{document}